\documentclass[10pt,letterpaper]{article}
\usepackage{opex3}

\usepackage{color} 
\usepackage{ulem}

\usepackage{cite}

\begin{document}

\title{Chromatic control in coextruded layered polymer microlenses}

\author{Michael Crescimanno,$^{1*}$ Tom N. Oder,$^{1}$  James H. Andrews,$^{1}$ Chuanhong Zhou,$^{1}$  Joshua B. Petrus,$^{1}$ Cory Merlo,$^1$ Cameron Bagheri,$^1$ Connor Hetzel,$^1$ James Tancabel,$^1$  Kenneth D. Singer,$^{2,3}$ and  Eric Baer$^{3}$} 

\address{$^1$  Dept. of Physics \& Astronomy, Youngstown State Univ., Youngstown, OH 44555, USA\\
$^2$ Dept. of Physics, Case Western Reserve Univ., Cleveland, OH 44106, USA\\
$^3$ Macromolecular Science and Engineering, Case Western Reserve Univ., Cleveland, OH 44106, USA}

\email{$^*$mjcrescimanno@ysu.edu}

\begin{abstract} 
We describe the formation, characterization and theoretical understanding of 
microlenses comprised of  alternating polystyrene  and polymethylmethacrylate layers produced by multilayer coextrusion. These lenses
are fabricated by  photolithography,  
using a grayscale mask followed by plasma
 etching, so that the refractive index alternation of the bilayer stack 
appears across the radius of the  microlens. The alternating quarter-wave thick layers form a one-dimensional photonic crystal whose dispersion augments the material dispersion, allowing one to sculpt the chromatic dispersion of the lens by adjusting the layered structure.
Using Huygen's principle, we model our 
experimental measurements of the focal length of these lenses across 
the reflection band of the multilayer polymer film from which the microlens
is fashioned. For a 56 $\mu$m diameter multilayered lens
of focal length 300 $\mu$m, we  measured a $\sim$ 25\% variation in the 
focal length across a shallow, 50 nm-wide reflection band. 
\end{abstract}



\section{Introduction}
\label{sec:intro}

Microlenses and microlens arrays play increasingly significant roles in miniaturizing optical systems for beam shaping and homogenization \cite{16}, interconnections, imaging, and displays.  
These arrays involve patterning of micron and submicron structures on functionalized and structured materials, with a rich history of success based on the invention of a wide variety of patterning techniques such as holographic photolithography \cite{10}, e-beam lithography \cite{11}, hot embossing/injection molding \cite{39, 40}, and imprinting \cite{22} or microcontact printing \cite{14}  with or without using solvents. A review of top-down and bottom-up patterning techniques applied to polymers can be found in  \cite{9}.  Popular methods of fabricating microlenses include the use of thermal reflow \cite{oder03, daly91}, micro droplets \cite{17}, and imprint molding \cite{22}.   
In addition, the use of grayscale photolithography has received recent attention due to several of its advantages. For example, 
the use of grayscale avoids the misalignment issues that arise from the use of multiple masks. Grayscale photolithography is less susceptible to distortion \cite{20} than soft- and nano-lithography imprint techniques.
Further, as described below, the use of grayscale lithography enables axial cutting and thus axial gradient index microlenses, allowing new applications not possible with a radial gradient index profile obtained by imprint molding.  Use of axially-cut multilayers may enable better control over the step size in a diffractive lens so that light diffracted from increasing ring diameters etched into the lens travels the same fixed distance to the focal point. (See \cite{19}.) 

Patterning using a grayscale mask followed by plasma etching to transfer the pattern to the substrate has been used for diffractive and refractive microoptics for beam shaping, spectral separation, phase modulation, lensing and other applications \cite{18, 41}. The mask used in grayscale photolithography is itself made using high energy beam sensitive (HEBS) glass, with the pattern being traced by electron-beam lithography. The exposure to high energy electron beams causes the reduction of silver ions in the glass. Areas of the mask exposed to a higher dose from the electron beam will have higher concentrations of reduced silver ions resulting in increased optical density (OD) there. By adjusting the grayscale levels in the pattern, the curvature can be controlled so that in a single exposure, a plano-convex photoresist shape can be obtained, circumventing exposure to heat used in a thermal reflow or a hot embossing process, which may otherwise be detrimental to some polymers. By this technique, ``axial'' gradient index microlenses can be fabricated whereby the index variation is along the lens axis and also appears on the lens sag. This is in contrast to ``radial'' gradient index where the index variation is radial outward from the center of the lens and can be fabricated by imprint molding. In the axial gradient microlenses, the index variation along the lens sag can be employed to precisely tailor the angle of refraction in order to provide superior corrections of different types of aberrations \cite{42,43}.  

At the same time,  layered polymer stacks formed by co-extrusion \cite{Ed} are being developed for a wide variety of optics applications, such as filters \cite{4}, sensors \cite{5}, switches and optical limiters \cite{6}, data storage media \cite{7}, and lasers \cite{8}.  These multilayer materials are attractive for their ready functionalization, ease of processing and amenability to large-area, low-cost fabrication \cite{2,laserreview,3}.   Most of these applications involve quarter-wave thick alternating layers of different refractive index forming a one-dimensional photonic crystal with a characteristic reflection band that depends on the  layer thicknesses, refractive index, and number of layers~\cite{kazmi07.01}.  Thus, the optical dispersion of the lens  is augmented by the dispersion characteristic of the layered photonic crystal medium.  This provides the ability, through careful choice of materials and structure, to sculpt the chromatic response  of the microlens elements.  
  
We report here on  a grayscale mask-based etching technique applied to a multilayer polymer stack to demonstrate the technical feasibility of making microlenses with large, designed chromatic dispersion. Beyond the process of fabricating these lenses, we characterize the focusing and spectral dispersion properties of an axially-terminated microlens array etched from the multilayer polymer film possessing a shallow reflection band in the visible. A straightforward physical optics  model qualitatively captures the salient features of these unique optical elements and may be useful for designing more complex layered lenses. 

\section{Materials and methods}
\label{sec:mnm}

The transparent constituent polymers used consisted of bilayers of polystyrene (PS) with an average refractive index of 1.6 and polymethylmethacrylate (PMMA) with an average refractive index of 1.5  made from a melt-process co-extrusion technique described in detail in  \cite{kazmi07.01}. The lens material had 32 bilayers, 64 individual layers in total, each layer about 80 nm thick, with an average layer-to-layer thickness variation of 22\%. The total thickness of the polymer material was about 10 $\mu$m. About 2 cm $\times$ 2 cm piece of the polymer sample was mounted on a clean silicon substrate for support and the top protective polyethylene layer was removed. A drop of thick photoresist (AZ P4620) was applied on the sample which was then spun at 2,500 rpm for 30 seconds on a Brewer Cee 200 spin coater, and baked at 90$^\circ$C for 1 minute on a hot plate. Note that this baking temperature is below the $T_g$ for PS (95$^\circ$C), PMMA ($\sim$100$^\circ$C) and AZ P4620 (125$^\circ$C). The sample was then exposed to UV light through the grayscale mask (described below) for 30 seconds using a Karl Suss MJB3 mask aligner and the sample was continuously swirled for about 2 minutes in a diluted AZ 400K developer solution, followed by rinsing in de-ionized (DI) water. 

\begin{figure}[tbp!]
\centering\includegraphics[scale=.85]{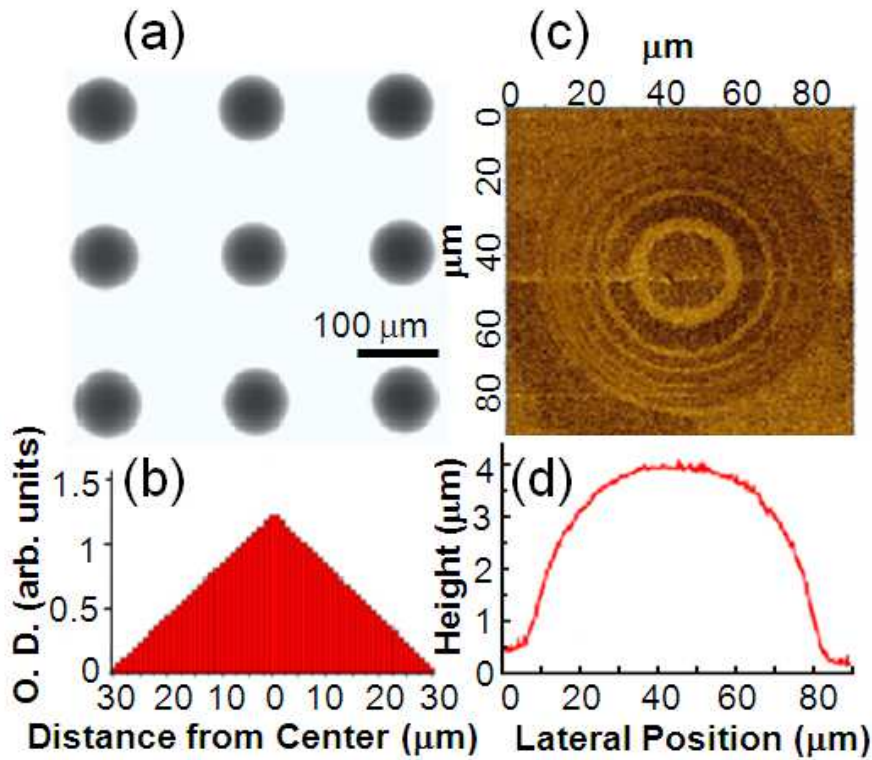}
\caption{(a)  Microscope image of grayscale mask (top left) with (b) radial optical density (O.D) profile (bottom left).  (c) Polymer multilayer microlens AFM image viewed in  contact mode reveals the salient layering of the different polymers (top right) and (d) a typical polymer multilayer microlens AFM cross-section (bottom right). }
\label{fig:lens_rings}
\end{figure}

The grayscale glass mask \cite{vendor} was designed such that, for a single microlens in the array, the optical density (OD) decreased radially outward through several gray levels from a value of 1.2 at the center to 0 at the edge so that the intensity of the UV light passing through the mask is varied accordingly. Figure \ref{fig:lens_rings}(a) shows the microscope image of the grayscale mask in which one can observe the variation in the shade across the circular cross-section, representing the variation of the OD in the circular structure. Each gray level was defined by a 1.0 $\mu$m wide ring so that for a 60 $\mu$m diameter microlens, there were 30 gray levels whose OD values decreased from 1.2 at the center by 0.04, as illustrated in the optical density profile shown in Fig. \ref{fig:lens_rings}(b). As further detailed by Canyon, Inc \cite{vendor}, the linearly varied OD values across the 1.0 $\mu$m-wide rings as shown in Fig.  \ref{fig:lens_rings}(c) can produce a smooth convex surface profile due to inherent proximity effects in the lithographic method and depending on the exposure parameters.  Thus, the resulting photoresist pattern is convex in shape following developing in the AZ 400K developer solution. The single exposure using the grayscale mask produces a convex shape without melting the AZ P4620 photoresist which would have required a temperature that could have ruined the multilayer stacks in the polymer material.  


Successful transfer of the plano-convex resist to the underlying multilayered polymer requires careful optimization of the etch parameters to remove both the photoresist and the polymer at about an equal rate. In our process, the optimum condition consisted of using a mixture of $O_2$ and $CF_4$ flowing at 2.0 and 3.0 standard cubic centimeters per minute at standard temperature and pressure; a total pressure of 100 mTorr and a radio frequency power of 20 watts. At these settings, an etch duration of 25 minutes succeeded in removing all the photoresist, thereby transferring the resist's post-exposure plano-convex shape into the multilayered polymer.

The lenses are made in square arrays that have a center-to-center separation distance of about 180 $\mu$m. Arrays of microlenses of four different sizes with lens diameters of nominally 80, 60, 40 and 20 $\mu$m were fabricated. The experimental data described here are from the 60 $\mu$m and 80 $\mu$m diameter microlenses. 

Figure \ref{fig:lens_rings}(c) is the atomic force microscope image (AFM Agilent Model 5500, friction in contact mode),  which shows that the resulting lens is cut from the multilayered polymer with a ring pattern that is determined by the (quasi-uniform) layer thicknesses and the lens shape in relief. The ring pattern was similarly observed in SEM images (variable pressure scanning electron microscope imaging using the JEOL, model JSM-IT300LV, not shown) of these microlens arrays. Figure  \ref{fig:lens_rings}(d) is the 2-dimensional AFM scan across a microlens showing that this particular example was etched to a depth of about 3.5 microns.

\section{Optical characterization}
\label{sec:exp}

\begin{figure}[tbp!]
\centering\includegraphics[scale=1]{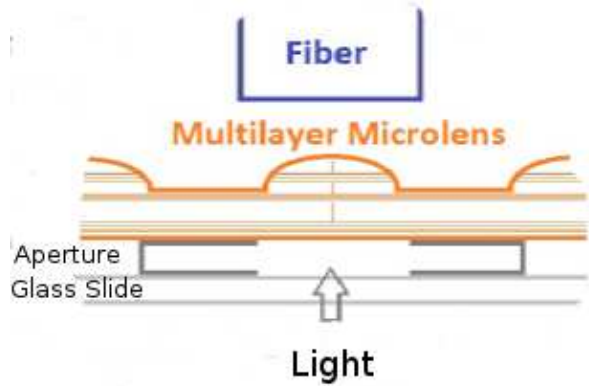}
\caption{Experimental setup for characterizing the spectral dispersion of the focusing properties of the microlenses.}
\label{fig:setup} 
\end{figure}

Our experimental set up for measuring the focal dispersion of a microlens is shown in Fig. \ref{fig:setup}. 
The particular microlens in the array to be tested is illuminated from behind by a broadband 
quartz-halogen source through an aperture ($\sim$ 200 $\mu$m diameter) 
mounted just behind the polymer multilayer. A 50-$\mu$m 
diameter optical fiber is mounted on a two-axis translation stage that enables the  fiber 
entrance face to be  hand-aligned with the symmetry axis of the microlens.  The microlens/aperture
combination is mounted on an electrically-actuated stage (on a kinematic mount) in order to enable more careful 
alignment between the microlens and the  optical fiber placed above it.  
Proper alignment is confirmed by
 the shape of the spectral output from the fiber as a function of distance from the microlens.

\begin{figure}[tbp!]
\centering\includegraphics[scale=.68]{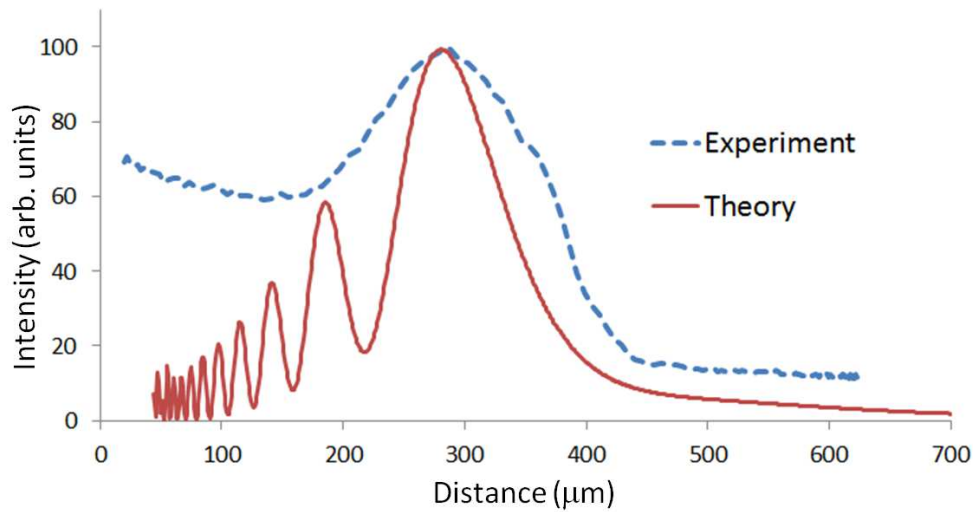}
\caption{ The dashed blue  trace is a typical multilayered co-extruded polymer microlens light curve of captured focused light. The wavelength for this light curve is 695$\pm$.5 nm and the focal length is $\sim$ 300 microns.  The solid red trace is an  idealized  light curve, from physical optics theory 
assuming monochromatic light ($\sim$ 695 nm) incident along the optical axis of
the microlens whose dimensions are given in Section 4.}
\label{fig:lightcurve} 
\end{figure}

A 60-second ramp applied to the actuating stage then withdraws the 
microlens longitudinally from contact with the collecting fiber to a distance 
of up to 1 mm. During this motion, every 0.2 seconds 
the spectrometer (an OceanOptics USB4000 UVVIS, 0.3 nm resolution) 
records the spectrum. Repeated fine adjustments to the alignment of the microlens with the fiber are made while scanning 
along the optical axis
through the focal length until the signal from the fiber  peaks as it passes axially 
through the focal point of the microlens.  
Once optimally aligned, about 300  spectra are recorded during 
each of multiple ramp cycles. These spectra are then reorganized by 
individual wavelength, yielding an intensity versus distance dataset (we
call a ``light curve'') for each wavelength. A representative measured 
light curve (dashed blue trace) at a single wavelength is shown in 
Fig. \ref{fig:lightcurve} along with a typical theory curve (solid red trace). 
The location of the peak indicates the approximate focal length.  
The qualitative difference between the theory and experiment is primarily a 
consequence of the non-ideality of the lens and underlying multilayer structure as well as the 
geometry/acceptance of the fiber.   
The calibration of the distances in the actuating 
stage is accomplished with a micrometer, yielding a distance versus current
calibration that is accurate to about 10\% in determining the absolute focal length, but can fix the dispersion much more precisely.

In addition, the focused beam waist was measured under white light using a NIKON Eclipse M600 microscope (objective 100$\times$/0.9) to determine the spot size on a 1.3 MPixels CMOS array and found to be $3.3 \pm 0.5 \mu m$, which is  close to the approximate expected diffraction limit of the microlens ($\sim$3$\mu m$).

\section{Theory and data analysis}
\label{sec:theory}

To model the dispersive optical properties of the multilayered microlens, rather than full Finite-Difference Time-Domain (FDTD) simulation, we 
 use Huygen's principle and add up the spherical waves coming from each 
section of the lens, while paying careful attention to the phase retardation at each surface
due to the wave's propagation through the 
 layered microlens \cite{bornwolf}. 

\begin{figure}[tbp!]
\centering\includegraphics[scale=1]{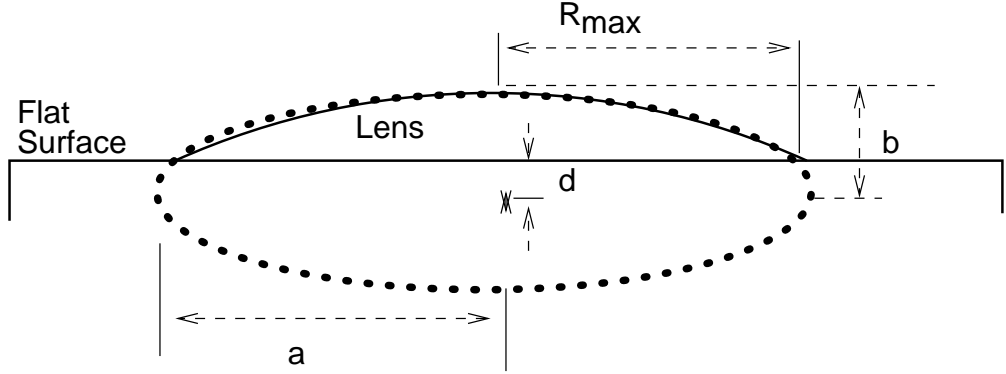}
\caption{Fixing the notation for a section of an oblate spheroid lens, showing ellipse semimajor (a) and semiminor (b) axes, the outer radius of the raised plano-convex lens shape $R_{max}$, and the depth of the center of the ellipse (d) below the plane of the surface supporting the lens. }
\label{fig:lens_geometry}
\end{figure}

Refer to Fig.  \ref{fig:lens_geometry} to fix the geometrical parameters we 
use to describe the lenses ({\it i.e.} as a section of an oblate spheroid formed by rotating an ellipse about its minor axis).   In terms of these parameters, the radius of curvature at 
the center of the lens is ${\tilde R} \sim {{a^2}\over{b}}$, and so, were it a monolithic lens of material index $n$, 
 the thin lens approximate focal distance would be $f\sim {{a^2}\over{(b-d)(n-1)}}$. Note that 
$R_{max} = a\sqrt{1-{{d^2}\over{b^2}} } $, in terms of which
 \begin{equation}
{\tilde R} = {{R_{max}^2}\over{b-d}} {{b}\over{b+d}},
\end{equation} 
an expression that smoothly interpolates between the elliptical and 
parabolic cross-section limits, in which ${\tilde R} = {{R_{max}^2}\over{2h}}$ 
and $f = {{\tilde R}\over{(n-1)}}$  and altitude $h = b-d$. By altitude we 
are referring to the maximum height of the microlens surface above the flat 
segment between lenses. 

Because we are most interested in the far-field, we simply add up spherical waves 
\begin{equation}
{\rm d} \psi = z_h {{e^{ik_0 r}}\over{r}}{\rm d} A 
\end{equation}
emanating from each surface ${\rm d} A$ of the ring of 
height $h$ making up the lens, with $z_h$ being the magnitude and phase of the wave as it emerges 
from that surface ($r$ is the distance to the point at which we are combining all of the waves) and
 where $k_0$ is the magnitude of the vacuum wavevector. Technically, the integration must include
 all of the illuminated surfaces, including the flat parts, but in the case of uniform illumination we
 can use linearity to  reduce it to just an integral across the lens. 

The final simplification we employ comes from the paraxial limit. In general, the computation of the far fields can be written as sums of elliptics or (in the case of azimuthal symmetry) 
Bessel functions \cite{bornwolf}.  In the paraxial limit, a distance 
$X>>2R_{max}$, but 
much closer to the optical axis than $X\lambda/R_{max}$, 
the integrations simplify and can be written in terms of 
exponentials. On axis, each annular section (of $r_{outer}$ outer
and $r_{inner}$ inner radius) at height $h$ of the lens contributes to 
the wave sum an amplitude $S_h$ given by 
\begin{equation}
S_h= {{1}\over{X-h}}\int_{r_{inner}}^{r_{outer}} ~~ r{\rm d}r e^{i{{k_0 r^2}\over{2(X-h)}}}(z_he^{-ik_0h}-1), 
\end{equation}
where we 
have removed an irrelevant overall phase  factor 
and have taken the difference between the amplitude and phase of the wave 
emerging from that height of material and that amplitude (unity) 
and phase ($e^{ik_oh}$)  that would have emerged from there {\it were there no lens there at all}. 

Thus, for each annulus that makes up the lens, the contribution it makes to the on-axis far field (at position $X$) is 

\begin{equation} 
S_h = {\cal A} (z_he^{-ik_0h}-1)[e^{i{{k_0 r_{outer}^2}\over{2(X-h)}}}-e^{i{{k_0 r_{inner}^2}\over{2(X-h)}}} ],
\label{each_annulus} 
\end{equation} 
where ${\cal A}$ is some overall amplitude which scales with the input wave's amplitude and is the same for all 
annuli. Thus, the total wave amplitude $S$ at a distance $X$ from the lens but on axis 
is $S_{tot} = \sum_h S_h$. These expressions are complicated to 
evaluate for layered materials 
because $z_h$ is an intricate function of $h$.  For the well known case of a 
monolithic plano-convex lens, however, one can compute this sum analytically, 
recovering in the thin lens limit the familiar results of a Gaussian beam focal 
distance and waist. 

To find $z_h$ rigorously would require that one solve Maxwell's equations for
the wave field in this layered geometry, or obtain it numerically via a 
finite element (for example, via FDTD) simulation. More intuitively, we 
adopt an approximate $z_h$, namely, we estimate $z_h$ from the 
transmission amplitude for this normal incident wave to 
exit a stack of these multilayers that would have 
terminated at $h$. Physically this approximation 
includes only internal reflections in the full multilayer 
from the layers below height $h$, discounting reflections from the rest of the 
lens that contribute to the wave emanating from the annulus at height $h$.  

The modulus square 
of the resulting $S_{tot}$ as a function of  $X$, the distance 
along the axis, typically leads to a light curve as in Fig. \ref{fig:lightcurve} (red solid trace) as compared with the measured light curve (blue dashed trace) for one of the layered microlenses.  
Note that the theory trace can be thought of  as plot of the intensity
 (essentially, $|S_{tot}|^2$) on the lens axis versus distance ($X$) from the lens. 
We identify the maximum of this 
light curve as the focal length of the lens at this wavelength, 
and note that at large $X$, ignoring background light, the intensity falls off 
universally as $1/X^2$, as expected. 
In experiment,
the interference fringes at intermediate $X$ are generally smeared out due to
spectral and spatial averaging. 

\begin{figure}[tbp!]
\centering\includegraphics[scale=.85]{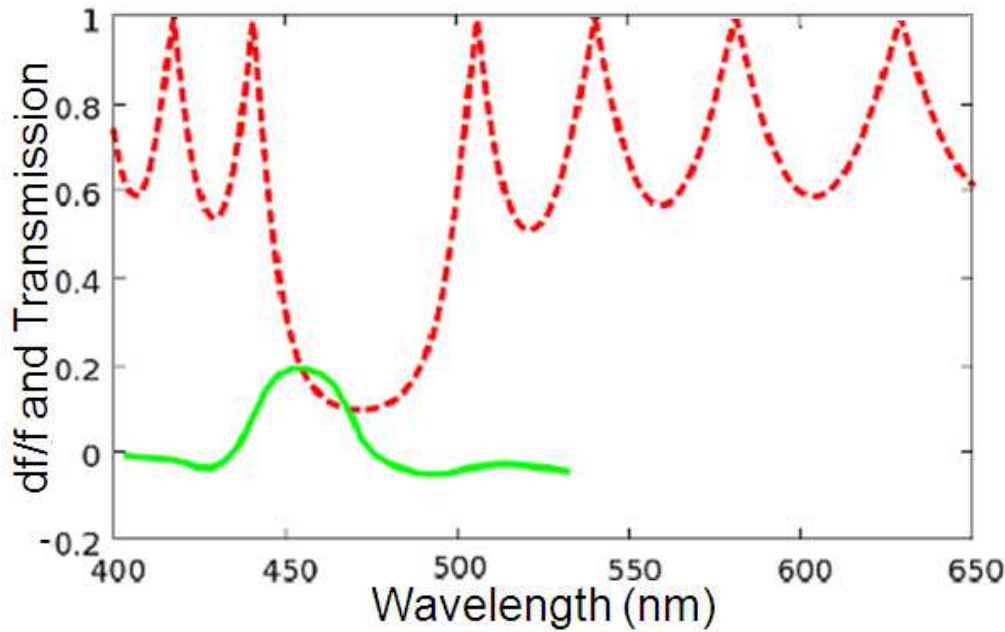}
\caption{Multilayered microlens computed focal length dispersion using physical optics theory described in the text. The dashed red trace is the computed transmission band of the 32 perfectly 
uniform layer model (layer indexes of 1.58 and 1.44) showing a  reflection band between 440 and 510 nm. In solid green is the computed fractional change in the focal length (df/f) of the multilayer structure showing a profound change around the reflection band edge. }
\label{fig:computed_focal}
\end{figure}

Equating the location of the  on-axis light curve maximum with the focal 
length, theory indicates that these lenses will typically have large, 
designable chromatic aberration. 
The solid green trace of Fig. \ref{fig:computed_focal} is a typical 
theoretical curve depicting $\delta f/f$, the 
fractional variation in the focal distance with wavelength, 
for a microlens with (compare Fig. \ref{fig:lens_rings}(d) and Fig. \ref{fig:lens_geometry}) $d=0.27$ microns, $a=34$ microns, $b=2.5$  microns  built from 
a 32-layer multilayer lens blank having a 
reflection band stretching from 430 nm to 510 nm (as shown in transmission curve (dashed red) of Fig. \ref{fig:computed_focal}). In essentially all cases,
the theory model predicts
 focal lengths for these multi-layered convex lenses that are shorter on the red (long wavelength)
side of the band edge than the blue side, opposite to the 
usual expectations of chromatic aberration in a solid lens of a material 
with normal dispersion. As described below, this shape can be understood in terms of the effects of the transmission band structure created by the multilayering (again, for example as shown by the dashed red trace in Fig. \ref{fig:computed_focal}). 
FDTD calculations for a plano-convex shape in Fig. \ref{fig:lens_geometry} reveal that there is little noticeable change in the 
focus spot size across the reflection band, and, further, that the spot size corresponds closely to that of a monolithic lens. As an 
additional check we note that these more detailed numerical calculations show the same qualitative changes in the focal length 
as the wavelength varies across the reflection band as are seen in the simple semi-analytical model described above (Fig. \ref{fig:computed_focal}) and 
as noted in experiment (Fig. \ref{fig:lens1}).

\section{Discussion}
\label{sec:disc}

\begin{figure}[tbp!]
\centering\includegraphics[scale=.85]{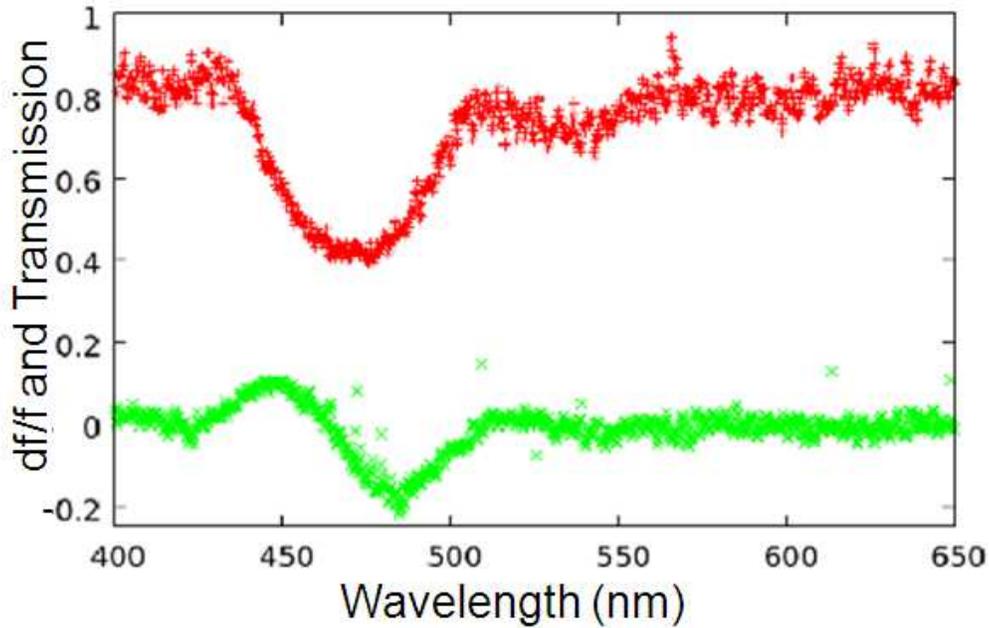}
\caption{Measured layered microlens focal length dispersion (green `$\times$') 
using the fiber transport system described in the text and the layered 
material's transmission spectrum (red `$+$'). The lens material
has an easily discernable shallow
reflection band from about 430 nm to 510 nm. The len's 
focal length dispersion is pronounced across the band, 
amounting to nearly 25\% changes in the
$\sim$300 $\mu$m focal length.}
\label{fig:lens1}
\end{figure}

Figure \ref{fig:lens1} shows the measured transmission 
band  and focal 
length variations of a single  multilayered PS/PMMA polymer microlens 
of radius 
$\sim$ 28$\pm$4 microns and height about 2.5$\pm$.5 microns. 
These measurements were made using AFM and the 
stated variations 
are due to process variations across the microlens array, not 
dimensional uncertainty 
in the measurements of a single lens. 
In that figure, the shallow 
reflection band stretching from 430 nm to 510 nm is easily discernable, as is 
the pronounced chromatic aberration in the focal distance of 
nearly 30 percent in a span of 25 nm. 
 The measured 
average focal length was $\sim$ 300$\pm$20 microns.  
Note that this is broadly consistent with expectations for a thin lens
of material whose index is the average of the constituent polymer indices.  
The scatter in the focal lengths at each wavelength 
is technical noise associated
with the jitter in the numerics associated with determining the maxima of each 
of the experimental light curves. Fitting the portion 
of that $df/f$ versus $\lambda$ curve
away from the reflection band reveals the nominal 
changes in the focal length expected from dispersion in these bulk
materials, albeit reduced somewhat from the diffractive effects from the 
smallness of the lens itself. For comparison, the 
expected (fractional, normal) chromatic aberration in a focal distance 
of a lens made from a (unlayered) PS/PMMA blend 
in this wavelength range (500-1000 microns) is expected to be roughly 2\% per octave, and a linear fit in this wavelength range of our data is consistent 
with roughly 1\% per octave. This and the regular 
shape of the light curves away from the material's reflection band (Fig.
\ref{fig:lightcurve}, blue dashed trace) further 
support the identification of this feature with a microlens of the shape 
imaged in Fig. \ref{fig:lens_rings}.

The qualitative similarity between the experimentally measured focal length 
dispersion and that derived from the simple physical optics  model can 
be readily understood in terms of where the light lingers as it flows through 
the material. Recall that for these binary multilayers, 
at the short wavelength edge of an optical band,
the light's electric fields are primarily in the low index material whereas, 
at the long wavelength band edge, they reside primarily in the high index 
material. Thus, because the effective index that the light fields respond to 
at the long wavelength side is closer to the lower index material's, the focal 
length is larger, and vice versa. This simple narrative also leads to an 
estimate for the upper bound of the focal length variation; because the 
focal length is inversely proportional to $(n-1)$, we expect a change across 
the band of $(n_{high}-n_{low})/({\bar n}-1) \sim 20\%$ (for a PS/PMMA layered
film), akin to that measured.   Further discussion of errors due to layer thickness variations and 
imperfections due to the photolithography etching process  are beyond the scope of this demonstration~\cite{20,dawson}.

\section{Conclusions and future work}
\label{sec:conc}

We have created layered microlenses with designed chromatic aberration 
in the visible. A simple physical optics model for the transmission and focusing of the 
light traversing these microlenses shows how the `structural' dispersion
caused by the layering leads to this designed chromatism.  
Even though such chromatism occurs in the multilayer material's reflection band,  the number and properties of the layers can be designed 
and tuned to still allow a substantial percentage of the light through, while preserving the chromatism. We have studied these microlens structures primarily as a test of our overall understanding of the 
role of designed  dispersion in layered microlenses, with potential applications in multi-spectral imaging, and dispersion correction and control. 

 Our work here builds upon earlier work demonstrating the benefits of a roll-to-roll co-extrusion process for multilayering to create Bragg reflectors~\cite{kazmi07.01,26} and distributed feedback lasers~\cite{8,laserreview,andre12.01}.    We emphasize that the simple multilayer stack design used is not necessarily ideal for applications and we did not investigate these materials for imaging applications, but we have used this system to better understand the feasibility of the novel combination of an axially-cut, lens-like pattern in a multilayer polymer.

Beyond the polymer combinations used here, this work connects the designed optical dispersion in metamaterials in the visible with the properties of a lens.  Depending upon the parameters of the etching process, the resulting shape may be smoothly varying through the layers or can be etched in a terraced layer-cake structure by taking advantage of different etching speeds in the different polymers.  In an effort to understand more fully the connection between multilayer design and the chromatic properties of the lens, we are currently trying to make multilayered lenses with higher numerical aperture as well as exploring novel polymer multilayer optical materials such as those with designed phase-slip defects formerly used in laser and magneto-optics 
studies \cite{26, andre12.01}. 
Optical transport in these phase-slip defect multilayers is well understood
and will make lenses ideal for more 
rigorous testing of our model.  
Because we can also 
create multi-band, gradient, and `chirped' structures as described in  \cite{gradientlayers}, we are exploring the utility of lenses made from materials with
more intricate customizable dispersion. 
For future development, we note that a grayscale mask can be used with multilayers to create a circular grating structure that has been shown to be an attractive structure for confining modes in a surface-emitting photonic bandgap laser~\cite{27}. Combining the surface structuring with the multilayer Bragg structure is
one approach to a more easily fabricated three-dimensional photonic crystal. 

\section*{Acknowledgments}
The authors are grateful to the National Science Foundation for financial support from the Science and Technology Center
for Layered Polymeric Systems under grant number No. DMR 0423914 and
separately under grant number ECCS-1360725 and also grant DMR 1229129.
We also acknowledge support from the State of Ohio, Department of Development, State of Ohio,  Chancellor of the Board of Regents and
Third Frontier Commission, which provided funding in support of the Research Cluster on Surfaces in Advanced Materials.


\begin{thebibliography}{10}
\newcommand{\enquote}[1]{``#1''}

\bibitem{16} S. Pfadler, F. Beyrau, M. L\"offler, and A. Leipertz, \enquote{{Application of a beam homogenizer to planar laser diagnostics},} Opt. Express {\bf 14}, 10171-10180 (2006).


\bibitem{10} A. Yariv and  M. Nakamura, \enquote{{Periodic structures for integrated optics},}  IEEE J. Quantum Electronics \textbf{13}, 233-253 (1977).

\bibitem{11} K.-S. Chen, I-K. Lin, and F.-H. Ko, \enquote{{Fabrication of 3D polymer microstructures using electron beam lithography and nanoimprinting technologies},}  J. Micromech. Microeng. {\bf 15}, 1894-1903 (2005).


\bibitem{39} R. F. Shyu and H. Yang, \enquote{{A promising thermal pressing used in fabricating microlens array},}  Int. J. Adv. Manuf. Technol. {\bf 36}, 53-59 (2008).

\bibitem{40} L. Peng, Y. Deng, P. Yi, and X. Lai, \enquote{{Micro hot embossing of thermoplastic polymers: a review},} J. Micromech. Microeng. {\bf 24}(1) 013001-1~-~013001-23 (2014).


\bibitem{22} Y. Xia, E. Kim, X.-M. Zhao, J.A. Rogers, M. Prentiss, and G.M. Whitesides, \enquote{{Complex optical surfaces formed by replica molding against elastomeric masters},}  Science {\bf 273}, 347-349 (1996).


\bibitem{14} E. Kim, Y. Xia, X.-M. Zhao, and G. M. Whitesides, \enquote{{Solvent-assisted microcontact molding: A convenient method for fabricating three-dimensional structures on surfaces of polymers},}  Adv. Mater. {\bf 9}, 651-654 (1997).

\bibitem{9} Z. Nie and E. Kumacheva, \enquote{{Patterning surfaces with functional polymers},} Nature Materials {\bf 7}, 277-290 (2008).


\bibitem{oder03}T. N. Oder, J. Shakya, J. Y. Lin, and H. X. Jiang, \enquote{{Nitride microlens arrays for blue and UV wavelength applications},} Appl. Phys. Lett. {\bf 82}(21), 3692-3694  (2003).
	

\bibitem{daly91} D. Daly, R. F. Stevens, M. C. Hutley, and N. Davies, in M. C. Hutley, ed., {\it Microlens Arrays}, IOP Short Meetings Series No. 30  (Institute of Physics,  1991), p. 23.

\bibitem{17} E. Bonaccurso, H.J.  Butt, B. Hankeln, B. Niesenhaus, and K. Graf,  \enquote{{Fabrication of microvessels and microlenses from polymers by solvent droplets},} Appl. Phys. Lett. {\bf 86}, 124101 (2005).

\bibitem{20} J. A. Rogers, K. Paul, and G. M. Whitesides, \enquote{{Quantifying distortions in soft lithography},} J. Vac. Sci. Technol. B {\bf 16}, 88-97 (1998).

\bibitem{19} S. A. Kemme and A. A. Cruz-Cabrera, \enquote{{Fabricating surface-relief diffractive optical elements},} in S. A. Kemme, ed., {\it Microoptics and Nanooptics Fabrication}   (CRC,  2010) pp. 1-38.

\bibitem{18} H.K. Wu,  T.W. Odom,  and G.M. Whitesides, \enquote{{Reduction photolithography using microlens arrays:Applications in gray scale photolithography},} Anal. Chem. {\bf 74}, 3267-3273 (2002).

\bibitem{41} J. J. Yang, Y. S. Liao, and C. F. Chen, \enquote{{Fabrication of long hexagonal micro-lens array by applying gray-scale lithography in micro-replication process},} Opt. Commun. {\bf 270}, 433-440 (2007). 

\bibitem{42} R. K. Wade, B. V. Hunter, B. Walters, and P. Fournier, \enquote{{Properties, specifications, and tolerances of GRADIUM glasses}.} 
in {\it  Optical Science, Engineering and Instrumentation '97}, (International Society for Optics and Photonics, 1997) pp. 63-74.

\bibitem{43} D. T . Moore, \enquote{{Gradient index optics}} in M. Bass, ed., {\it Handbook of Optics: Vol II}, (McGraw-Hill Professional, 2004), ch. 9. 

\bibitem{Ed}  A. C. Edrington, A. M. Urbas, P. DeRege,, C. X. Chen, T. M. Swager, N. Hadjichristidis, M. Xenidou, L. J. Fetters, J. D. Joannopoulos, Y. Fink, and E. L. Thomas, \enquote{{Polymer-based photonic crystals},}  Adv. Mater. \textbf{13}, 421-425 (2001).


\bibitem{4} M. Kimura, K. Okahara, and T. Miyamoto, \enquote{{Tunable multilayer-film distributed-Bragg-reflector filter},} J. Appl. Phys. \textbf{50}, 1222-1225 (1979).

\bibitem{5} J. Kunzelman, B. R. Crenshaw, and C. Weder, \enquote{{Self-assembly of chromogenic dyes - a new mechanism for humidity sensors},} J. Mater. Chem. \textbf{17}, 2989-2991 (2007).

\bibitem{6} J. W. Kang, E. Kim, and J. J. Kim, \enquote{{All-optical switch and modulator using photochromic dye doped polymer waveguides},} Opt. Mater. \textbf{21}, 543-548 (2002).

\bibitem{7} C. Ryan, C. W. Christenson, B. Valle, A. Saini, J. Lott, J. Johnson, D. Schiraldi, C. Weder, E. Baer, K.D. Singer, and J. Shan, \enquote{{Roll-to-roll fabrication of multilayer films for high capacity optical data storage},} Adv. Mater. \textbf{24}, 5222-5226 (2012).

\bibitem{8} H. Song, K. Singer, J. Lott, Y. Wu, J. Zhou, J. Andrews, E. Baer, A. Hiltner, and C. Weder, \enquote{{Continuous melt processing of all-polymer distributed feedback lasers},}  J. Mater. Chem. \textbf{19}, 7520-7524 (2009).

\bibitem{2} R. R. S\o ndergaard, M. H\"{o}sel, and F. C. Krebs, \enquote{{Roll-to-roll fabrication of large area functional organic materials},}  J. Polym. Sci., Part B: Polym. Phys.  \textbf{51}, 16-34 (2012).

\bibitem{laserreview} J. H. Andrews, M. Crescimanno, K. D. Singer, and E. Baer, \enquote {{Melt-processed polymer multilayer distributed feedback lasers: progress
and prospects},}  J. Polym. Sci., Part B: Polym. Phys. \textbf{52}, 251-271 (2014).


\bibitem{3} J. Hu, L. Li, H. Lin, P. Zhang, W. Zhou, and Z. Ma, \enquote{{Flexible integrated photonics: where materials, mechanics and optics meet},}  Opt. Mater. Express \textbf{3}, 1313-1331 (2013). 

\bibitem{kazmi07.01}
T.~Kazmierczak, H.~Song, A.~Hiltner, and E.~Baer, \enquote{{Polymeric
  one-dimensional photonic crystals by continuous coextrusion},}  Macromol.
  Rapid Commun. \textbf{28}, 2010-2016 (2007).

\bibitem{vendor} Fabricated by Canyon Materials, Inc. 6665 Nancy Ridge Drive, San Diego, CA 92121, USA. http://www.canyonmaterials.com/CMI-01-88-5.html 

\bibitem{bornwolf} M. Born and E. Wolf,  {\it Principles of Optics}, 6$^{th}$ Ed. (Cambridge University,  1980) pp. 435-441.

\bibitem{26} K. D. Singer, T. Kazmierczak, J. Lott, H. Song, Y. Wu, J. Andrews, E. Baer, A. Hiltner, and C. Weder, \enquote{{Melt-processed all-polymer distributed Bragg reflector laser},}  Opt. Express {\bf 16}, 10358-10363 (2008).

\bibitem{dawson} N. Dawson, K. D. Singer, J. H. Andrews, M. Crescimanno, G. Mao, J. Petrus, H. Song, and E. Baer, \enquote{{Post-
Process Tunability of Folded One-Dimensional All-Polymer Photonic Crystal Microcavity Lasers},}  Nonlin. Opt.
Quantum Opt. {\bf 45}, 101-111 (2012).

\bibitem{andre12.01}
J.~H. Andrews, M.~Crescimanno, N.~J. Dawson, G.~Mao, J.~B. Petrus, K.~D.
  Singer, E.~Baer, and H.~Song, \enquote{{Folding flexible co-extruded
  all-polymer multilayer distributed feedback films to control lasing},} 
 Opt. Express \textbf{20}, 15580--15588 (2012).

\bibitem{gradientlayers} M. Ponting, T. M. Burt, L. T. J. Korley, J.H. Andrews, A. Hiltner, and E. Baer, ``Gradient multilayer films by forced assembly coextrusion,''  Ind. Eng. Chem. Res. {\bf 49}(23), 12111-12118 (2010).

\bibitem{27} C. Bauer, H. Giessen, B. Schnabel, E.-B. Kley, C. Schmitt, U. Scherf, and R. F. Mahrt, \enquote{{A surface-emitting circular grating polymer laser},} Adv. Mater. {\bf 13}, 1161-1164 (2001).


\end{thebibliography}
\end{document}